\documentclass[aps, prl, superscriptaddress, preprint]{revtex4-1}

\usepackage{amsmath}
\usepackage{amsfonts}
\usepackage{amssymb}
\usepackage{graphicx}

\begin{document}

\title{Dynamic conductivity scaling in photoexcited V$_2$O$_3$  thin films}

\author{Elsa Abreu}
\affiliation{Department of Physics, Boston University, Boston, Massachusetts 02215, USA}
\email{elsabreu@bu.edu}

\author{Siming Wang}
\affiliation{Department of Physics, The University of California at San Diego, La Jolla, California 92093, USA}
\affiliation{Center for Advanced Nanoscience, The University of California at San Diego, La Jolla, California 92093, USA}
\affiliation{Materials Science and Engineering Program, The University of California at San Diego, La Jolla, California 92093, USA}

\author{Gabriel Ramirez}
\affiliation{Department of Physics, The University of California at San Diego, La Jolla, California 92093, USA}
\affiliation{Center for Advanced Nanoscience, The University of California at San Diego, La Jolla, California 92093, USA}

\author{Mengkun Liu}
\affiliation{Department of Physics, The University of California at San Diego, La Jolla, California 92093, USA}
\affiliation{Department of Physics, Stony Brook University, Stony Brook, New York 11790, USA}

\author{Jingdi Zhang}
\affiliation{Department of Physics, Boston University, Boston, Massachusetts 02215, USA}

\author{Kun Geng}
\affiliation{Department of Physics, Boston University, Boston, Massachusetts 02215, USA}

\author{Ivan K. Schuller}
\affiliation{Department of Physics, The University of California at San Diego, La Jolla, California 92093, USA}
\affiliation{Center for Advanced Nanoscience, The University of California at San Diego, La Jolla, California 92093, USA}
\affiliation{Materials Science and Engineering Program, The University of California at San Diego, La Jolla, California 92093, USA}

\author{Richard D. Averitt}
\affiliation{Department of Physics, Boston University, Boston, Massachusetts 02215, USA}
\affiliation{Department of Physics, The University of California at San Diego, La Jolla, California 92093, USA}
\email{raveritt@bu.edu}

\begin{abstract}
Optical-pump terahertz-probe spectroscopy is used to investigate ultrafast far-infrared conductivity dynamics during the insulator-to-metal transition (IMT) in vanadium sesquioxide (V$_2$O$_3$). The resultant conductivity increase occurs on a tens of ps timescale, exhibiting a strong dependence on the initial temperature and fluence. We have identified a scaling of the conductivity dynamics upon renormalizing the time axis with a simple power law ($\alpha \simeq 1/2$) that depends solely on the initial, final, and conductivity onset temperatures. Qualitative and quantitative considerations indicate that the dynamics arise from nucleation and growth of the metallic phase which can be described by the Avrami model. We show that the temporal scaling arises from spatial scaling of the growth of the metallic volume fraction, highlighting the self-similar nature of the dynamics.
Our results illustrate the important role played by mesoscopic effects in phase transition dynamics.
\end{abstract}

\maketitle


The variety of electronic, magnetic and structural phases exhibited by transition metal oxides arise from a delicate balance between competing degrees-of-freedom whose contribution to the macroscopic properties is challenging to ascertain \cite{Rondinelli2011}. An increasingly successful approach to tackle this problem is that of time resolved experiments, where ultrafast excitation and probing enables the determination of the fundamental timescales of the material down to femtosecond resolution \cite{Schuller1976, Orenstein2012, Zhang2014}. Access to specific energy scales and modes of the system is made possible by ultrafast sources ranging from terahertz (THz) to x-rays frequencies. Initial all-optical measurements of electron-phonon relaxation in metals \cite{Brorson1990, Sun1994} have paved the way to time resolved investigations of complex systems, from spins in magnetic materials \cite{vanKampen2002, Zhu2005, Kampfrath2011, Vicario2013, Graves2013} to superconducting gaps \cite{Demsar2003, Matsunaga2012} or to surface charges in topological insulators \cite{Wang2013}.

To date, most time-resolved experiments in complex transition metal oxides, and in particular the vanadates, have focused on microscopic dynamics \cite{Cavalleri2001, Cavalleri2004, Kubler2007, Mansart2010, Pashkin2011, Liu2011, Wall2012, Cocker2012}. For example, fast sub-ps electronic and structural responses have been reported for vanadium dioxide \cite{Cavalleri2001, Cavalleri2004, Kubler2007, Pashkin2011, Wall2012, Cocker2012}, though the precise driving mechanism of the IMT remains unclear \cite{Biermann2005, Lazarovits2010}. However, it is increasingly evident from static measurements that nano-to-meso scale phase coexistence plays a crucial role in determining the properties of complex materials, including cuprates, manganites, and vanadates \cite{Dagotto2005, Basov2011, Qazilbash2011, Liu2013}. This naturally extends to the dynamic investigation of the phase coexistence stage, as investigated in VO$_2$ \cite{Hilton2007, Pashkin2011, Liu2012, Cocker2012}. At a minimum, neglecting mesoscale effects can lead to a misinterpretation of the dynamics. More importantly, as shown in this letter, mesoscale dynamics are of intrinsic interest from fundamental and applied perspectives.

V$_2$O$_3$ is a paramagnetic metal with rhombohedral crystal symmetry \cite{McWhan1969, McWhan1971, McWhan1973, Qazilbash2008, Stewart2012, Hansmann2013} that undergoes a first order phase transition to an antiferromagnetic insulating state at $T_{IMT} = 175$K, accompanied by a change to a monoclinic crystal structure \cite{McWhan1970}.  In this letter, we present the mesoscopic conductivity dynamics of V$_2$O$_3$ across the insulator-to-metal transition following an optical initiated  picosecond thermal quench into the metallic state. Importantly, we identify scaling of the conductivity dynamics upon renormalizing the time axis with a simple power law ($\alpha \simeq 1/2$) that depends solely on the \textit{experimentally determined temperatures}. These temperatures consist of the initial temperature  $T_i$,  final temperature, $T_f$ (determined by $T_i$ and the incident fluence, $F_{inc}$), and conductivity onset temperature, $T_{0.5} = 160$K (defined as the onset of a macroscopic THz conductivity which, for a $2d$ system, occurs at a volume fraction of $f = 0.5$). Further, the temporal evolution of the conductivity is well fit by the Avrami model consistent with nucleation and growth of the metallic phase. In conjunction with scaling, this allows us to demonstrate that the temporal rescaling arises from spatial scaling of the metallic volume fraction. Thus, the mesoscopic conductivity dynamics of the IMT are dictated by a length scale, $R(t)$, associated with metallic phase domain coarsening. Our analysis further indicates that the growth of the metallic phase is ballistic, occurring at the sound velocity. \\


Growth of $95nm$ thick V$_2$O$_3$ films was performed in an ultrahigh purity Ar environment by rf magnetron sputtering of a V$_2$O$_3$ target onto an r-plane (10$\bar{1}$2) sapphire substrate \cite{Stewart2012}.
X-ray diffraction characterization indicates near single crystal growth following the substrate orientation. 
Transient conductivity measurements are performed using $1.5$eV, $50$fs pulses from a $3$mJ Ti:Sapph amplifier. Incident pump fluences, $0.5$ to $4$mJ/cm$^2$, remain below the damage threshold for V$_2$O$_3$ \cite{Liu2011}. THz probe pulses are generated and detected in $1mm$ thick ZnTe crystals.

\begin{figure} [htb]
\begin{center}
\includegraphics[width=0.5\textwidth,keepaspectratio=true]{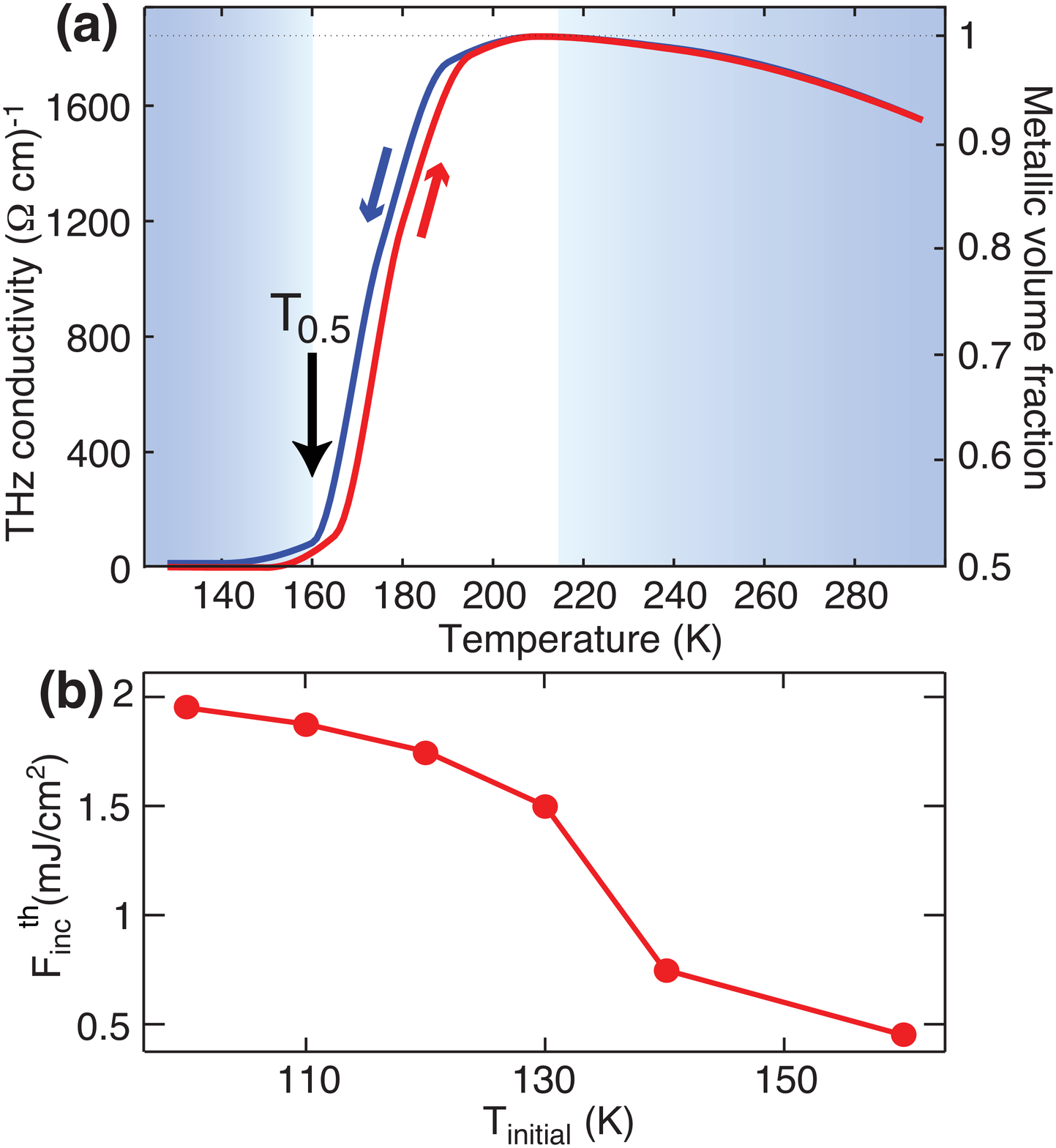}
\caption{a) V$_2$O$_3$ thin films characterization: $\sigma$ vs. $T$ from THz time-domain spectroscopy. The arrow marks the conductivity onset temperature, $T_{0.5}=160$K. $f(T)$, calculated from Eq. \ref{EMA} for the unshaded region of the plot, is shown on the right-hand side axis. b) Fluence threshold to dynamically drive a finite $\Delta \sigma(t)$, for $T_i<T_{0.5}$.}
\label{fig1}
\end{center}
\end{figure}

We first present the V$_2$O$_3$ static conductivity characterization using THz time-domain spectroscopy (i.e. the photoexcitation is blocked).  The temperature dependent real part of the Drude conductivity, $\sigma(T)$, is shown in Fig. \ref{fig1}(a). The IMT occurs at $T_{IMT} \simeq 175$K, with a narrow hysteresis associated with its first order nature \cite{Ramirez2009}. As indicated by the arrow in Fig. \ref{fig1}(a), $T_{0.5} = 160$K corresponds to the temperature above which a finite THz conductivity arises.

\begin{figure} [htb]
\begin{center}
\includegraphics[width=1\textwidth,keepaspectratio=true]{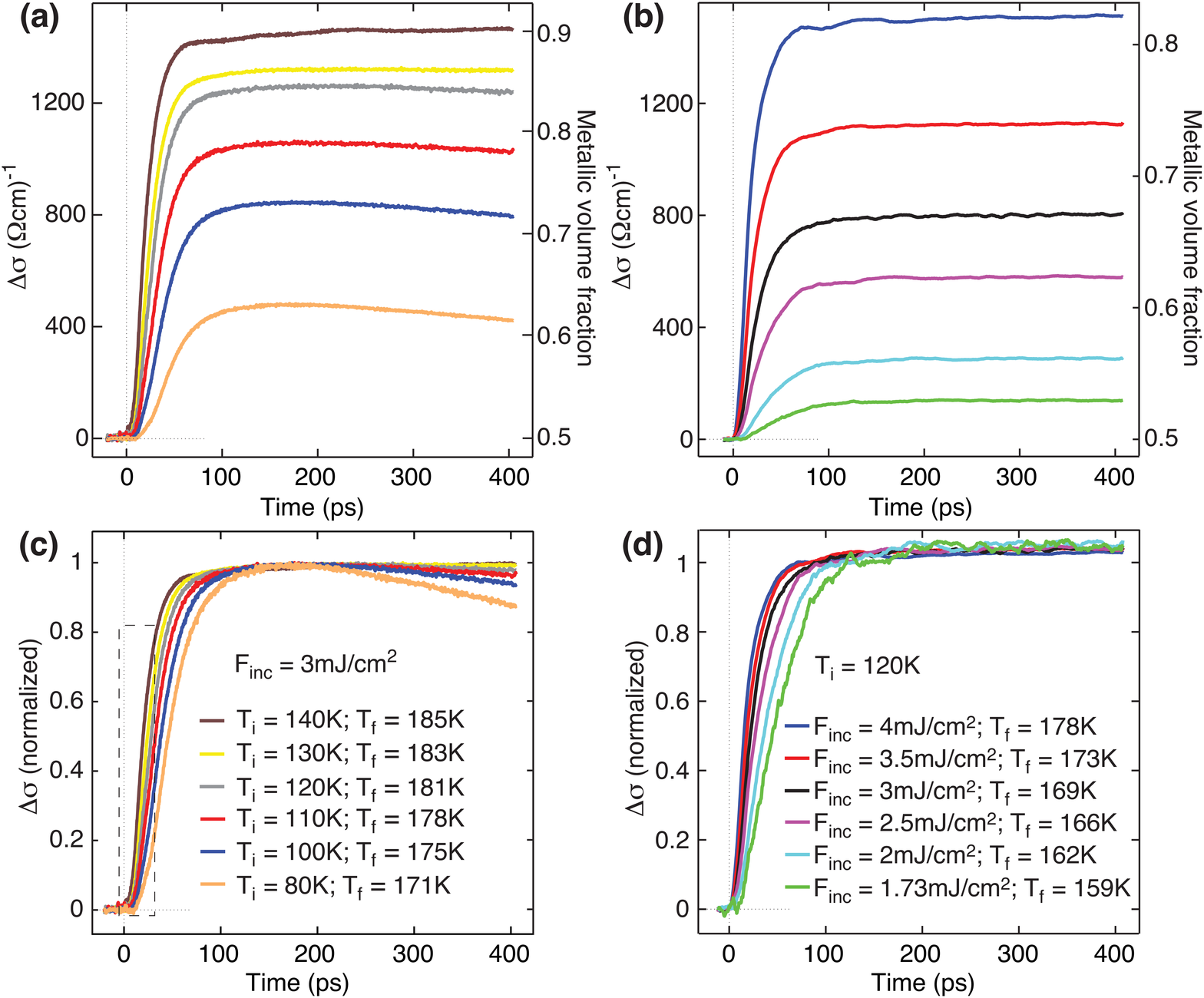}
\caption{THz conductivity dynamics of V$_2$O$_3$ thin films following a) a $3$mJ/cm$^2$ optical excitation, for $T_i<T_{0.5}$, and b) a $1.73 - 4$mJ/cm$^2$ optical excitation, for $T_i=120$K. c) and d): normalization of the data in a) and b), respectively, revealing the $T_i$ and $F_{inc}$ dependence of the $\Delta \sigma(t)$ rise time. A delayed onset of $\Delta \sigma(t)$ is visible for lower $T_i$ (the region within the dashed box of c) is magnified in Fig. \ref{fig4}(b)).}
\label{fig2}
\end{center}
\end{figure}

Optical-pump THz-probe experiments were performed for several $T_{i}$ ($<T_{0.5}$) and $F_{inc}$ values.
Photoexcitation at $1.5eV$ initiates an ultrafast heat quench in the V$_2$O$_3$ film, where excited electrons relax via phonon emission in $\sim 1$ps (as determined from the two temperature model \cite{Kaganov1957, Anisimov1974, Allen1987, Brorson1990}). In other words, heating from $T_{i}$ to $T_{f}$ occurs in $\sim 1$ps, setting up a nonequilibrium situation where the insulating phase is unstable, leading to metallic phase growth. As such, changes in the transient THz conductivity $\Delta \sigma(t)$ are reflective of nucleation and growth dynamics.
One of the hallmarks of a photoinduced phase transition is the observation of a fluence threshold, $F_{inc}^{th}$, for the onset of the IMT, as shown in Fig. \ref{fig1}(b) as a function of $T_i$. $F_{inc}^{th}$ decreases with increasing $T_i$  and is in line with what is observed in VO$_2$ \cite{Hilton2007, Pashkin2011}. The determination of $F_{inc}^{th}$ is made possible by detailed measurements of the conductivity dynamics which are considered in greater detail in Fig. \ref{fig2}.
Figure \ref{fig2}(a) shows $\Delta \sigma(t)$ (for various $T_{i}$) for $F_{inc} = 3$mJ/cm$^2$, corresponding to an absorbed energy density of $\sim 170$J/cm$^3$. $\sigma(t)$ increases over $10s$ of ps following photoexcitation, and saturates at long times to a value corresponding to $\sigma(T_{f})$. In fact, comparison of $\sigma(t=400$ps$,T_{f})$ with Fig. \ref{fig1}(a) provides a means to estimate $T_{f}$, and is consistent with two temperature model estimates (Fig. S1 \cite{supMaterial}). Varying $F_{inc}$ at fixed $T_i$ leads to variations in $T_f$ and consequently in $\Delta \sigma$, as shown in Fig. \ref{fig2}(b).
Clearly, the dynamics depend on both $T_i$ and $T_f$.

Further insight into the $\Delta \sigma(t)$ dynamics can be obtained by normalizing the data, as illustrated in Figs. \ref{fig2}(c) and \ref{fig2}(d). Fig. \ref{fig2}(c) shows that the $\Delta \sigma(t)$ rise time is faster for increasing $T_i$. For instance, for $T_{i} = 80$K the maximum in $\Delta \sigma(t)$ is reached in $\sim120$ps, while for $T_{i} = 140$K it takes $\sim60$ps.
The $\Delta \sigma$ rise time is also faster for increasing $F_{inc}$, as shown in Fig. \ref{fig2}(d).
A detailed analysis of these rise time dynamics will be presented below and constitutes the main result of this letter.

A partial recovery of $\Delta \sigma(t)$ is observable with decreasing $T_i$, as shown in Figs. \ref{fig2}(a) and \ref{fig2}(c).
Recovery on this timescale is unlikely due to heat escape from the sample which typically takes several nanoseconds \cite{Bechtel1975, Demsar1997, Wen2013}, though our data does not allow us to unequivocally rule out such a $100$ps-scale decrease of $\Delta \sigma(t)$ due to cooling.
The recovery may be related to the decreased stability of the metallic volume fraction distribution at low $T_f$, associated with a larger fraction of metallic regions whose characteristic dimensions are too small to undergo stable growth \cite{Caviglia2012}.\\

\begin{figure} [htb]
\begin{center}
\includegraphics[width=0.5\textwidth,keepaspectratio=true]{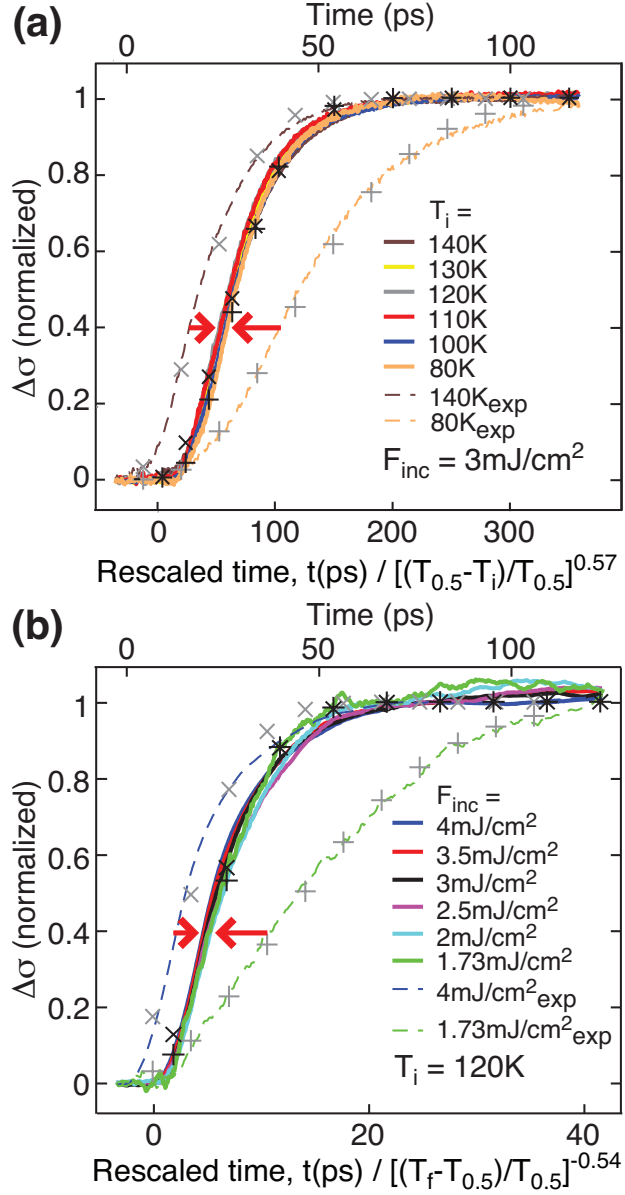}
\caption{Normalized conductivity dynamics for a) $F_{inc}=3$mJ/cm$^2$, with varying $T_i$, and for b) $T_i=120$K, with varying $F_{inc}$. The bottom (top) time axis corresponds to the scaled (unscaled) data, shown by full (dashed) lines. Note that the $t=0$ positions for the top and bottom time axes have been horizontally offset for clarity. Grey (black) crosses correspond to fits (scaled fits) to the a) $80$K and $140$K data, and to the b) $1.73$mJ/cm$^2$ and $4$mJ/cm$^2$ data, using Eq. \ref{Avrami} with $n=2$.}
\label{fig3}
\end{center}
\end{figure}

The qualitative discussion of the conductivity dynamics presented above suggests the primary role of nucleation and growth, with a clear dependence on $T_i$ and $T_f$. In the following, we investigate these dynamics in greater detail, first demonstrating their temperature dependent scaling. The temperature above which a macroscopic conductivity can be measured, $T_{0.5}$, is the critical temperature characteristic of the nucleation and growth process that underlies the IMT. It is therefore reasonable to expect a dependence of the IMT dynamic properties on $|T-T_{0.5}|$ \cite{KhomskiiBook2010}.
The $\Delta \sigma(t)$ curves collapse by scaling the time axis as shown in Fig. \ref{fig3}. In Fig. \ref{fig3}(a) $F_{inc}$ is kept fixed at $3$mJ/cm$^2$ and $T_i$ is varied between $80$K and $140$K (cf. Figs. \ref{fig2}(a) and \ref{fig2}(c)). Scaling of the time axis by the dimensionless factor, $t\rightarrow t/\Big(\frac{T_{0.5}-T_i}{T_{0.5}}\Big)^\alpha$, with $\alpha \simeq 1/2$, leads to a collapse of all the curves with different $T_i$ values (Fig. \ref{fig3}(a)). The same scaling leads to the collapse of the $\Delta\sigma$ dynamics for $F_{inc} = 2$mJ/cm$^2$ (Fig. S2 \cite{supMaterial}), which is close to $F_{inc}^{th}$ at low temperatures (Fig. \ref{fig1}(b)). However, varying $T_i$ corresponds to a variation not only of $T_i$ but also of $T_f$. A variation of $T_f$ alone can be achieved by fixing $T_i$ and varying $F_{inc}$. Fig. \ref{fig3}(b) shows results for a fixed $T_i=120$K and for $F_{inc}$ values between $1.73$ and $4$mJ/cm$^2$ (cf. Figs. \ref{fig2}(b) and \ref{fig2}(d)). A collapse of the normalized $\Delta \sigma(t)$ curves arises if the time axis is rescaled by $t\rightarrow t/\Big(\frac{T_f-T_{0.5}}{T_{0.5}}\Big)^{-\alpha}$ (Fig. \ref{fig3}(b)).
Notably, the scaling behavior relies on the experimentally determined $T_i$, $T_f$ and $T_{0.5}$ values. The only parameter that is varied to achieve the scaling shown in Fig. \ref{fig3} is the exponent $\alpha$.

To determine the value of $\alpha$  that provides the best scaling of the data, a scaling error was calculated for each value of $\alpha$ \cite{supMaterial}.
The optimal values of $\alpha$ are seen to lie close to $1/2$ and these minimized values were used to scale the data in Figs. \ref{fig3}(a) and \ref{fig3}(b). This temporal scaling is quite remarkable, indicating that the processes underlying the conductivity dynamics must also exhibit scaling. Further, the mean-field-like exponent of $1/2$ suggests that fluctuations are not dominant, and that a simple model can provide additional insights \cite{Binder1987, Chaikin2000}. In the following, we consider these results in terms of the nucleation and growth of the metallic volume fraction, $f(t)$.\\


The IMT in V$_2$O$_3$ is known to arise from nucleation and growth of metallic domains in an insulating background \cite{Lupi2010, Mansart2012, McLeod2014}.
The metallic volume fraction, $f(T)$, can be calculated from $\sigma(T)$ using the Bruggeman effective medium approximation:
\begin{equation}
f\frac{\sigma_m-\sigma}{\sigma_m+(d-1)\sigma}+(1-f)\frac{\sigma_i-\sigma}{\sigma_i+(d-1)\sigma}=0,
\label{EMA}
\end{equation}
where $\sigma_m$ and $\sigma_i$ correspond to metallic and insulating state conductivities, respectively, and the dimensionality $d=2$ for thin films \cite{Choi1996, Hilton2007}. In the temperature range where $\sigma(T)>>\sigma_i$, taking  $\sigma_i=0$ is a valid approximation, and Eq. \ref{EMA} yields a linear dependence of $\sigma(t)$ on $f(t)$, $\sigma(t)=\big(2f(t)-1\big)\sigma_m$. The right axis of Fig. \ref{fig1}(a) shows $f(T)$ across the IMT. The correspondence between $\sigma(T)$ and $f(T)$ values derived from THz time-domain spectroscopy data is strictly valid only in the $\sim 160 - 200$K range (unshaded region of the $\sigma(T)$ curve). For $T>200$K, $f(T)=1$ and the decrease in $\sigma(T)$, consistent with previous reports, arises from correlation effects, beyond a simple thermally induced increase in the scattering rate \cite{Qazilbash2008, Liu2011, Stewart2012, Hansmann2013}.
For $T<160$K, the $\sigma_i=0$ approximation in Eq. \ref{EMA} breaks down. DC resistivity measurements yield a thermally activated $\sigma_i(T)$, which we use to estimate $f(T)$ below $T_{0.5}=160$K from Eq. \ref{EMA}. It is clear that a non-zero $\sigma_i(T)$ for $T < T_{0.5}$ implies a non-zero $f(t)$ well below $T_{0.5}$. This is an important consideration for time-resolved experiments, where the initial condition is a mixed phase with metallic volume fraction $f(T_{i})$.

Classical models of nucleation and growth predict a dynamic evolution of the volume fraction $f(t)$, which can be described by the Avrami equation \cite{PTBook},
\begin{equation}
f(t) = 1-e^{-K \: t^n},
\label{Avrami}
\end{equation}
where $K$ is the rate at which $f(t)$ increases, and $n$ is an exponent that depends on the dimensionality and nature of the nucleation and growth.
As mentioned above, for $\sigma>>\sigma_i$, a linear relationship exists between $\sigma(t)$ and $f(t)$. The photoinduced conductivity variations we measure, $\Delta \sigma (t)$, are therefore proportional to $f(t)$, and Eq. \ref{Avrami} can be used to fit the normalized $\Delta \sigma (t)$ data.
A good fit is obtained for $n=2$ \cite{supMaterial}, as illustrated by the grey crosses in Figs. \ref{fig3}(a) and \ref{fig3}(b). 
Rescaling the time axis effectively corresponds to a rescaling of $K$. This is highlighted by the rescaled fitting curves, shown as black crosses in Figs. \ref{fig3}(a) and \ref{fig3}(b), which are obtained by replacing $K$ by $K /\Big(\frac{T_{0.5}-T_i}{T_{0.5}}\Big)^{n \alpha}$ (Fig. \ref{fig3}(a)) and by $K/\Big(\frac{T_f-T_{0.5}}{T_{0.5}}\Big)^{-n \alpha}$ (Fig. \ref{fig3}(b)) in Eq. \ref{Avrami}, while keeping the time axis unchanged. Such a behavior implies a temperature dependence of $K$, $K \propto 1/(T_{0.5}-T_i)$ and $K \propto (T_f-T_{0.5})$, i.e. the IMT is faster for increasing $T_i$ and $T_f$.

\begin{figure} [htb]
\begin{center}
\includegraphics[width=0.5\textwidth,keepaspectratio=true]{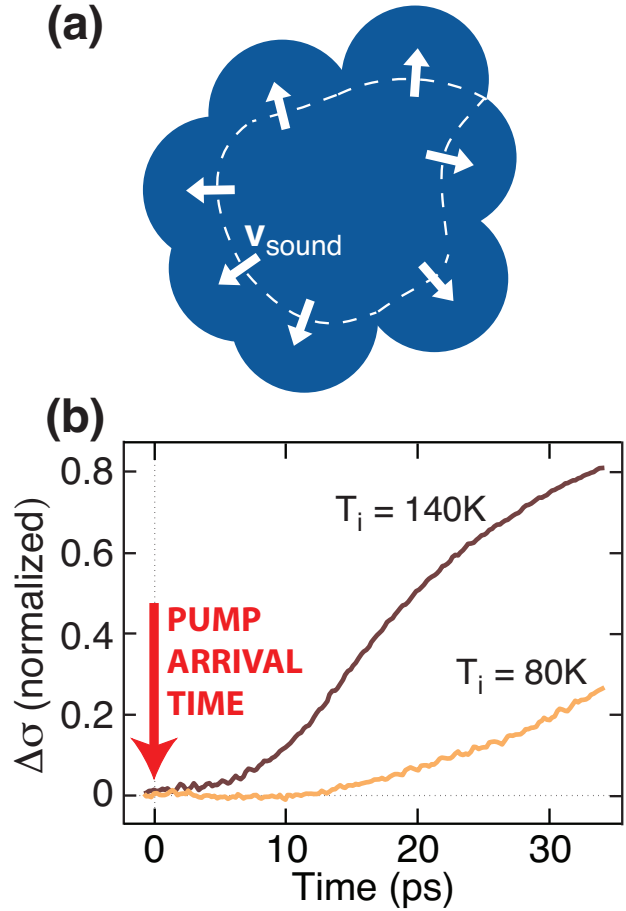}
\caption{a) Schematics of the $2d$ nucleation and growth process: growing (following photoexcitation) metallic regions are shown in blue, and the radial growth velocity, $\sim v_{sound}$, is indicated by the white arrows. b) Detailed view of the delayed $\Delta \sigma(t)$ onset from Fig. \ref{fig2}(c). No change in $\Delta \sigma(t)$ is observed during the first few picoseconds after the optical pump.}
\label{fig4}
\end{center}
\end{figure}

Both nucleation and growth contribute to $K$. In the current experiments, prior to the optically induced quench to $T_f$, the sample is at $T_i$ with a volume fraction of metallic nuclei $f_{i} = f(T_{i})$. The ultrafast quench to $T_f$ modifies the free energy landscape with a shift of the minimum from the insulating to the metallic phase.
Therefore, the energy gain associated with the IMT following photoexcitation drives the growth of metallic domains leading to an increasing volume fraction.
A model that is consistent with $n=2$ describes two-dimensional interfacial growth with quickly exhausted nucleation \cite{PTBook}. This yields
\begin{equation}
K = \frac{\pi}{2} \rho v^{2},
\label{K}
\end{equation}
where $\rho$ is the domain density and $v$ is the growth velocity. A schematic of this process (in $2d$) is shown in Fig. \ref{fig4}(a), with the metallic regions, shown in blue, growing at velocity $v$, as indicated by the white arrows. Eq. \ref{K} contains independent contributions from growth, through $v$, and nucleation, through $\rho$, allowing for additional insight into the experimental conductivity dynamics. We note that in thin film samples excited homogeneously across the entire thickness (the penetration depth of the optical pump is on the order of the film thickness) the growth is essentially $2d$ (in-plane growth) \cite{Choi1996}. 

We first consider the growth of the metallic regions where the interface (domain wall) separating the metallic and insulating phases propagates at $v$, defining a characteristic size given by the local radius of curvature $R(t)=v\times t$. It is worth noting that the structural transition which accompanies the IMT in V$_2$O$_3$ implies that $v$ cannot exceed the propagation velocity for structural distortions, i.e. the sound velocity, $v_{sound}$. Growth of the metallic phase at $v_{sound}$ is ballistic rather than diffusive, and is in line  with previous descriptions of thermally driven IMT in vanadates as martensitic, i.e. diffusionless \cite{Shadrin2000, Lopez2002}. An estimate of $K$ using $v_{sound}$ for V$_2$O$_3$ strongly suggests that growth is in the ballistic limit. Making the assumption that $\rho$ is given by the effective nuclei density $ \rho= 5\times10^{12} m^{-2}$ \cite{Lopez2002} and taking $v_{sound} = 7\times10^3m/s$ \cite{Seikh2006} yields $K = 3.8\times10^{-4} $ps$^{-2}$, consistent with the $30 \sim 70$ps rise times we observe (corresponding to $K^{-1/2}$, from Eq. \ref{Avrami}). Of course the assumption that $\rho$ corresponds to the equilibrium defect density is rather crude, but nonetheless verifies the applicability of the Avrami model and suggests interfacial ballistic growth of the metallic phase as the underlying dynamics of the macroscopic conductivity increase following photoexcitation.

Given that the sound velocity in V$_2$O$_3$ varies little with temperature \cite{Seikh2006}, the temperature dependence (and hence the scaling) of $K$ arises from nucleation processes that give rise to $\rho$. That is, in the $T_i$ and $T_f$ range we are exploring in these experiments, the metallic domain density $\rho$ must exhibit $\rho \propto 1/(T_{0.5}-T_i)$ and $\rho \propto (T_f-T_{0.5})$. Since the growth is ballistic (i.e. it proceeds at the maximum allowed velocity), it is changes in $\rho$ that determine the conductivity rise time dynamics. This means that for increasing $T_i$, the initial domain density available for growth increases. Additionally, with increasing $T_f$, $\rho$ increases meaning that the photoinduced thermal quench increases the domain density available for subsequent growth. Thus, the evolution of $\rho$ is fairly complicated and strongly dependent on the details of the nucleation process.

Nucleation has been reported to occur preferentially at defect sites on VO$_2$ and V$_2$O$_3$ \cite{Hansmann2013}. Such defect pinning effects can influence the IMT dynamics, leading to two limiting situations. For samples with a large defect density the nucleation is entirely heterogeneous and is expected to occur quasi-instantaneously (e.g. $\sim 1$ps), so that all nuclei are immediately available for growth \cite{PTBook, Lopez2002}. In clean samples, on the other hand, nucleation is essentially homogeneous and new nuclei formed during a finite period after the quench constitute a sizable fraction of the overall nuclei density from which the growth proceeds \cite{PTBook, Rethfeld2002}.
Neither of these limits appear to be completely verified in our experiments.

Importantly, the time delay, $\Delta t$, for the onset of $\Delta \sigma(t)$ (when $f(t)=0.5$) following photoexcitation is quite long, on the order of a few picoseconds (Figs. \ref{fig2} and \ref{fig4}(b)). $\Delta t$ is longer for lower $T_i$, as detailed in Fig. \ref{fig4}(b), where for $T_{i} = 80$K, $\Delta t>10$ps, while for $T_{i} = 140$K, $\Delta t\sim5$ps. The longer $\Delta t$ for lower $T_i$ is consistent with a smaller $f(T_{i})$. We have observed that $\Delta t$ is sample dependent hinting at the possibility of variations of the nucleation process with defect density, such that films with a lower extrinsic defect density have a larger homogeneous nucleation contribution.
However, homogeneous nucleation models predict an exponential increase of the nucleation rate (not to be confused with $K$) with superheating ($T_f > T_{0.5}$), i.e. with $T_f - T_{0.5}$ \cite{supMaterial, PTBook, Rethfeld2002}, while in the heterogeneous limit the dependence on $T_f-T_{0.5}$ becomes a power law \cite{supMaterial, Lopez2002}. Thus, the experimentally observed power law scaling attests to a significant heterogeneous contribution to the nucleation. Our analysis indicates that the dynamics can, to a certain extent, be controlled by the defect density. The higher the defect density, the faster the mesoscopic conductivity will be established after photoexcitation. This would come at the expense of a reduced metallic state conductivity.

More detail will be achievable once samples with controllable defect densities become available \cite{Ramirez2014}, since our data suggests that a change in the $T_f$ dependence of $K$ is to be expected as nucleation becomes more homogeneous (i.e. decreased defect density).
A change in the initial time delay, $\Delta t$ (Fig. \ref{fig4}(b)), would also be expected with variations in defect density. That is, with heterogeneous dominated nucleation, $\Delta t$ would be shorter since growth could proceed immediately, whereas for increasingly homogeneous nucleation, $\Delta t$ would increase since nuclei would need to be formed prior to growth. There is some evidence of this in our studies of different samples.
Details of the scaling exponent $\alpha$ should be accessible by Monte Carlo simulations.
Further insight into $\alpha$ can also be gained from considering different system dimensionalities \cite{supMaterial}. As discussed in Ref. \cite{supMaterial}, the geometry of domain growth is sensitive to the dimensionality of the system, and so is $n$ in Eq. \ref{Avrami}. The dynamic scaling we identified provides a simple yet robust means to analyze nucleation and growth dynamics during first order transitions.
In general, from an experimental perspective, an obvious follow-up to the present work would be to analyze samples with different morphologies and a controlled defect density, and using time resolved techniques that reveal the spatial distribution of the metallic domains at the mesoscale.

In summary, our conductivity dynamics investigations of the IMT in V$_2$O$_3$ thin films reveal the temperature dependence of domain growth through dynamic scaling of the $\Delta \sigma (t)$ rise time. These results highlight the importance of the mesoscale in shaping the dynamic evolution of first order IMTs. Such a temperature dependence of the dynamics provides additional control over the properties of transition metal oxides. This dependence must, furthermore, be taken into account when investigating materials where phase coexistence plays a significant role in the IMT.

\newpage

\begin{acknowledgments}
The authors would like to thank A. Polkovnikov and A. Sandvik for useful discussions. E.A. and R.D.A. acknowledge support from DOE - Basic Energy Sciences under Grant No. DE-FG02-09ER46643. E.A. acknowledges support from Funda\c c\~ao para a Ci\^encia e a Tecnologia, Portugal, through doctoral degree fellowship SFRH/ BD/ 47847/ 2008. The research at UCSD (S.W., G.R., I.K.S.) was supported by the AFOSR Grant No. FA9550-12-1-0381.
\end{acknowledgments}

\bibliographystyle{apsrev4-1}
\bibliography{V2O3Scaling}

\newpage

\end{document}


\title{Dynamic conductivity scaling in photoexcited V$_2$O$_3$  thin films\\
Supplemental Material}

\author{Elsa Abreu}
\affiliation{Department of Physics, Boston University, Boston, Massachusetts 02215, USA}
\email{elsabreu@bu.edu}

\author{Siming Wang}
\affiliation{Department of Physics, The University of California at San Diego, La Jolla, California 92093, USA}
\affiliation{Center for Advanced Nanoscience, The University of California at San Diego, La Jolla, California 92093, USA}
\affiliation{Materials Science and Engineering Program, The University of California at San Diego, La Jolla, California 92093, USA}

\author{Gabriel Ramirez}
\affiliation{Department of Physics, The University of California at San Diego, La Jolla, California 92093, USA}
\affiliation{Center for Advanced Nanoscience, The University of California at San Diego, La Jolla, California 92093, USA}

\author{Mengkun Liu}
\affiliation{Department of Physics, The University of California at San Diego, La Jolla, California 92093, USA}
\affiliation{Department of Physics, Stony Brook University, Stony Brook, New York 11790, USA}

\author{Jingdi Zhang}
\affiliation{Department of Physics, Boston University, Boston, Massachusetts 02215, USA}

\author{Kun Geng}
\affiliation{Department of Physics, Boston University, Boston, Massachusetts 02215, USA}

\author{Ivan K. Schuller}
\affiliation{Department of Physics, The University of California at San Diego, La Jolla, California 92093, USA}
\affiliation{Center for Advanced Nanoscience, The University of California at San Diego, La Jolla, California 92093, USA}
\affiliation{Materials Science and Engineering Program, The University of California at San Diego, La Jolla, California 92093, USA}

\author{Richard D. Averitt}
\affiliation{Department of Physics, Boston University, Boston, Massachusetts 02215, USA}
\affiliation{Department of Physics, The University of California at San Diego, La Jolla, California 92093, USA}

\maketitle

\section{1. Two temperature model}
\begin{figure} [htb]
\begin{center}
\includegraphics[width=0.5\textwidth,keepaspectratio=true]{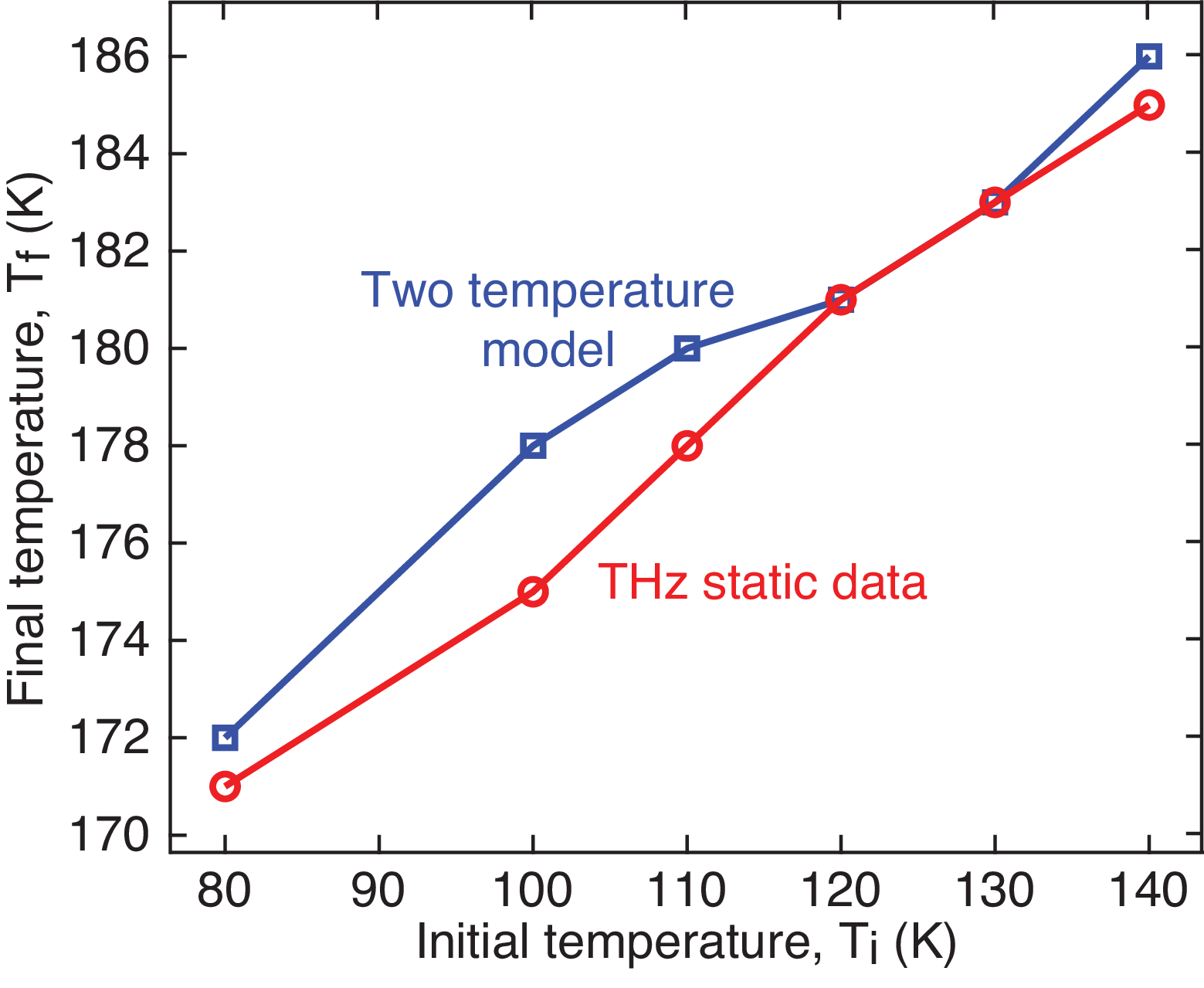}
\caption{Comparison between $T_f$ values predicted from the two-temperature model (blue, squares) and those obtained from the THz time-domain spectroscopy data (red, circles). Values are shown for the data taken with a $3mJ/cm^2$ fluence, with varying $T_i$.}
\label{figS1}
\end{center}
\end{figure}

The two-temperature model \cite{Kaganov1957, Anisimov1974, Allen1987, Brorson1990} was used to estimate the photoinduced sample heating, taking into account the properties of V$_2$O$_3$ \cite{Keer1976, Stewart2012, Qazilbash2008} and the incident pump fluence. $T_f$ predictions from the model are consistent with the values obtained by comparing the photoinduced conductivity change with the static $\sigma$ vs. $T$ curve of Fig. 1(a). This consistency is illustrated in Fig. S\ref{figS1}, for data taken with a $3mJ/cm^2$ fluence, with varying $T_i$. The photoinduced $\Delta \sigma$ discussed in this work is therefore seen to be caused by ultrafast heating of the sample.

\section{2. Scaling}

\subsection{2.1 $2mJ/cm^2$ data}
\begin{figure} [htb]
\begin{center}
\includegraphics[width=0.5\textwidth,keepaspectratio=true]{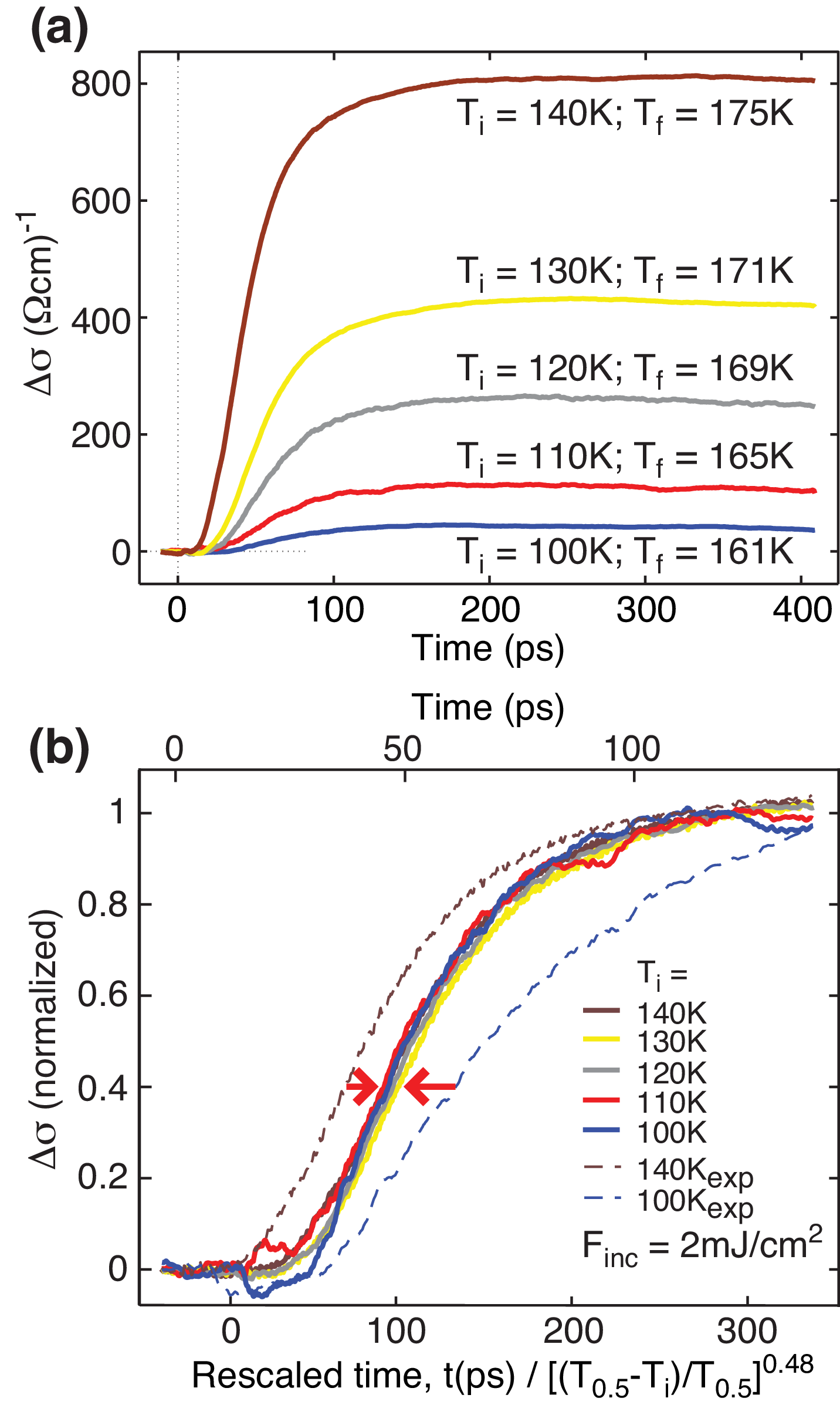}
\caption{a) THz conductivity dynamics of V$_2$O$_3$ thin films following a $F_{inc}=2mJ/cm^2$ optical excitation, for varying $T_i<T_{0.5}$. b) Normalization of the data in a). The bottom (top) time axis corresponds to the scaled (unscaled) data, shown by full (dashed) lines. Note that the $t=0$ positions for the top and bottom time axes have been horizontally offset for clarity.}
\label{figS2}
\end{center}
\end{figure}

Fig. S\ref{figS2} shows $\Delta \sigma (t)$ data for $2mJ/cm^2$, as a complement to the data shown in Figs. 2 and 3. Fig. S\ref{figS2}(a) shows the conductivity dynamics, as measured with optical-pump THz-probe. Fig. S\ref{figS2}(b) shows the normalized and rescaled data. Detailed discussion of these results is given in the main text.

\subsection{2.2 Scaling error calculation}
To determine the value of $\alpha$  that provides the best scaling of the data, a scaling error, $e_{scaling}(\alpha)$, was calculated for each value of $\alpha$.
$e_{scaling}(\alpha)$ is defined as
\begin{equation*}
e_{scaling}(\alpha) = \frac{\displaystyle{\sum_{\sigma_N = 0.05}^{\sigma_N = 0.95}(t_S(\sigma_N) - t_F(\sigma_N))^2}}{(t_S(0.95)-t_F(0.05))^2},
\end{equation*}
where $\sigma_N$ is the normalized $\sigma$ value, and $t_S$ ($t_F$) is the pump-probe delay time corresponding to $\sigma_N$ for the unscaled data set with slowest (fastest) $\tau$. For example, for the data in Fig. 3(a) $t_S(\sigma_N)$ ($t_F(\sigma_N)$) uses the $80$K ($140$K) data.  $\sigma_N = 0.05$ and $\sigma_N = 0.95$ were chosen as limiting values in order to avoid the interference of data noise, which would become a concern closer to $\sigma_N = 0$ and $\sigma_N = 1$. The normalization factor in the denominator allows for a comparison between rescaled data sets with different time axes.

\begin{figure} [htb]
\begin{center}
\includegraphics[width=0.5\textwidth,keepaspectratio=true]{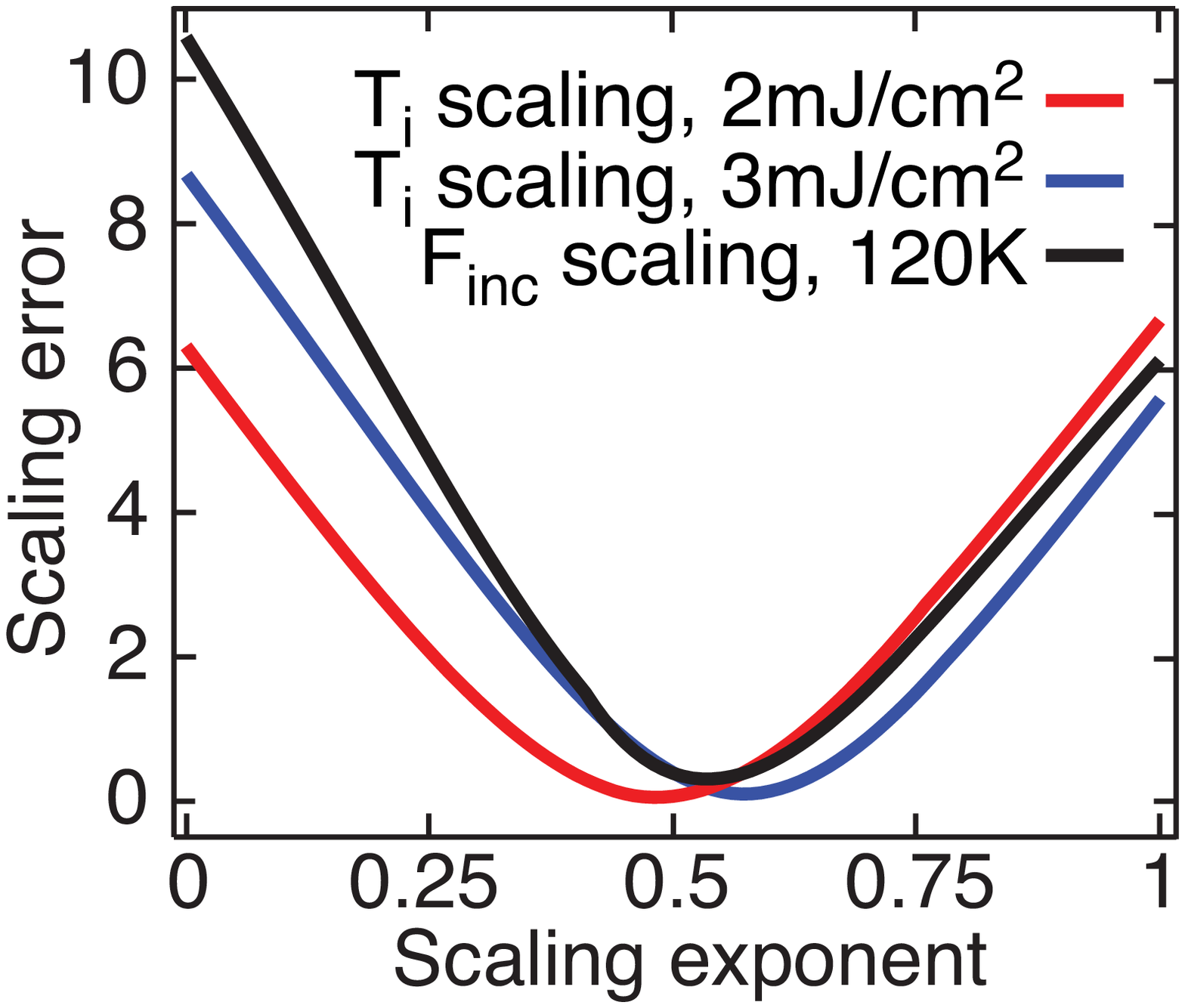}
\caption{Scaling error vs. scaling exponent, for the data in Figs. 3(a), 3(b) and S\ref{figS1}(b).}
\label{figS3}
\end{center}
\end{figure}

Fig. S\ref{figS3} plots $e_{scaling}(\alpha)$ for each of the data sets in Figs. 3(a), 3(b) and S\ref{figS2}. The optimal values of $\alpha$ are seen to lie close to $1/2$.\\

\section{3. Nucleation and growth models}

\subsection{3.1 Fits to the Avrami model}
The THz time-domain spectroscopy probe is only sensitive to $\sigma$ values corresponding to $f>0.5$. Therefore, the present data strictly only account for the dynamics leading from $f=0.5$ to the final metallic volume fraction determined by $T_f$, $f_f$, i.e. to the transition dynamics in a portion of the sample volume.
Fits to the Avrami equation (Eq. 2) were performed for the finite $\Delta \sigma (t)$ values measured with the THz time-domain spectroscopy probe. Due to the delay $\Delta t$, mentioned in the main text, $t-\Delta t$ was used for the fits, instead of simply $t$, leading to the fitting function: $f(t) = 1-e^{-K \: (t-\Delta t)^n}$.

\subsection{3.2 Homogeneous vs. heterogeneous nucleation}
For heterogeneous nucleation all nucleation sites are immediately available at $t\sim 1ps$.
If, on the other hand, nucleation has some homogeneous contribution, then new nuclei formed after the quench constitute a sizable fraction of the overall nuclei density from which growth proceeds \cite{Rethfeld2002}. In this case, in order to maintain $n=2$ as well as $2d$ ballistic growth, nucleation must be quickly exhausted, implying a nucleation rate $I(t) = I_0/t$ rather than $I(t) = I_0$, with $I_0$ constant. The ultrafast nature of the photoinduced heat quench lends support to such a quickly exhausted nucleation model.

In the case of homogeneous nucleation, $I_0$ is given by \cite{PTBook}:
\begin{equation*}
I_0 \propto e^{-c_1\frac{T_{0.5}^2}{T (T-T_{0.5})^2}}e^{-c_2\frac{1}{T}},
\end{equation*}
where $c_1$ and $c_2$ are constants. This dependence of $I_0$ on $T$ highlights the fact that superheating (i.e. $T_f > T_{0.5}$) increases the nucleation rate, so that an ultrafast quench to $T=T_f>T_{0.5}$, following photoexcitation, leads to more rapid IMT dynamics.
For increasingly heterogeneous nucleation the dependence of $I_0$ on $T_f$ deviates from the equation above, transitioning to a power law behavior \cite{Lopez2002}.

\subsection{3.3 Avrami exponent n=2 for homogeneous vs. heterogeneous nucleation}
The derivation of the Avrami equation for the volume fraction, $f(t) = 1- e^{-f_{exp}(t)}$, requires the expression for the expanded volume fraction, $f_{exp}(t)$. $f_{exp}(t)$ corresponds to the volume fraction that would be accessible if growth of a nucleus were not sterically hindered by neighboring domains, and $f_{exp}(t)>1$ is therefore possible \cite{PTBook}.
In the heterogeneous nucleation limit, when all nucleation sites are quasi-instantaneously available, only growth imparts a time dependence to $f_{exp}(t)$, and $f_{exp}(t) \propto t^2$ for $2d$ ballistic growth.
For homogeneous nucleation the time dependence of $I$ must be taken into account. Assuming a quickly exhausted model, i.e. $I(t) = I_0/t$, and $2d$ ballistic growth, the determination of $f_{exp}(t)$ requires integration of $I(\tau)(t-\tau)^2 d\tau$, also leading to $f_{exp}(t) \propto t^2$.
In conclusion, both nucleation scenarios imply an $n=2$ Avrami exponent, as stated in the main text.\\

\section{4. Scaling exponents for one- and three-dimensional growth}

The data presented in the main text corresponds to the case of $2d$ ballistic growth in thin films, depicted in Fig. 4(a), for which $n=2$ (Eq. 2).
In systems with different geometries growth occurs essentially in $1d$, as in nanobeams, or $3d$, as in bulk samples.
As alluded to in the main text, as system dimensionality varies, so does the Avrami exponent, so that $n=1$ for $1d$ and $n=3$ for $3d$ \cite{PTBook}.
Exponent values relative to the scaling with $T_i$, $\alpha_{T_i}$, and with $T_f$, $\alpha_{T_f}$, are expected to vary with dimensionality, following the variation of $\rho$, as detailed in the following paragraphs (the notation follows that in Eqs. 2 and 3 of the main text).

\subsection{4.1 Scaling with $T_i$}
As mentioned in the main text, $\rho^{2d} \propto 1/(T_{0.5}-T_i)^{n\alpha_{T_i}^{2d}} \simeq 1/(T_{0.5}-T_i)$ for $n=2$ and $\alpha_{T_i}^{2d} \simeq 1/2$.
$\rho^{2d}$ is the $2d$ nuclei density, given by the ratio of the number of nuclei to the $2d$ volume (or area), $\rho^{2d} = N_{2d} / V_{2d}$. It is reasonable to assume that $N_{1d} = N_{2d}^{1/2} = N_{3d}^{1/3}$, and obviously $V_{1d} = V_{2d}^{1/2} = V_{3d}^{1/3}$, so that $\rho^{1d} = (\rho^{2d})^{1/2} = (\rho^{3d})^{1/3}$.
Therefore, we can predict $\rho^{1d} \propto 1/(T_{0.5}-T_i)^{1/2} \simeq 1/(T_{0.5}-T_i)^{n\alpha_{T_i}^{1d}}$, i.e. $\alpha_{T_i}^{1d} = 1/2$ since $n=1$.
Similarly, $\rho^{3d} \propto 1/(T_{0.5}-T_i)^{3/2} \simeq 1/(T_{0.5}-T_i)^{n\alpha_{T_i}^{3d}}$, i.e. $\alpha_{T_i}^{3d} = 1/2$ since $n=3$.

\subsection{4.2 Scaling with $T_f$}
The scaling with $T_f$ arises due to an increased nucleation rate with superheating, leading to an increased nuclei density so that $\rho \propto (T_f-T_{0.5})^{n\alpha_{T_f}^{2d}} \simeq T_f-T_{0.5}$ for $n=2$ and $\alpha_{T_f}^{2d} \simeq 1/2$.
The nucleation rate is intrinsically microscopic and can be thought of as a nucleation probability for available nucleation sites, so that the variation with $T_f$ should be independent of the dimensionality of the system (and of its mesoscale growth). In this case, $\rho \propto (T_f-T_{0.5})^{n\alpha_{T_f}} \simeq T_f-T_{0.5}$ is valid for $1d$, $2d$ and $3d$. Since $n$ varies with dimensionality, we expect $\alpha_{T_f}^{1d} = 1$ and $\alpha_{T_f}^{3d} = 1/3$.

\subsection{4.3 Heterogeneous vs. homogeneous nucleation}
As mentioned in the main text, the dynamics we observe suggest a nucleation model with a strong heterogeneous character but likely some homogeneous contribution.
For the specific ratio of heterogeneous vs. homogeneous nucleation that characterizes our V$_2$O$_3$ films, $\alpha_{T_i}=1/2$ and $\alpha_{T_f}=1/2$. However, $\alpha_{T_f}$ might vary with varying ratio of heterogeneous vs. homogeneous nucleation, though our data does not allow us predict its exact evolution. In the homogeneous nucleation limit the dependence of $\rho$ on $T_f-T_{0.5}$ becomes exponential and $\alpha_{T_f}$ no longer applies. $\alpha_{T_i} \rightarrow 0$ is expected in the same limit, where $\rho \rightarrow 0$ for $T_f < T_{0.5}$, so that a decreasing $\alpha_{T_i}$ with increasingly homogeneous nucleation is expected.

\newpage

\bibliographystyle{apsrev4-1}
\bibliography{V2O3ScalingArxiv}

\newpage

